%% file: arxiv.tex
\documentclass[journal]{IEEEtran}
\usepackage{amsmath,amsfonts}
\usepackage{algorithmic}
\usepackage{algorithm}
\usepackage{array}
\usepackage{textcomp}
\usepackage{stfloats}
\usepackage{url}
\usepackage{verbatim}
\usepackage{graphicx}
\usepackage{cite}
\usepackage{amsmath,amsthm}   
\usepackage{amssymb}
\usepackage{nomencl}
\usepackage{multirow}

\usepackage{bm}
\usepackage{lipsum}
\usepackage{makeidx}
\usepackage{enumerate}
\usepackage{color}

\usepackage[font=small,labelfont=bf]{caption}
\usepackage[font=small,labelfont=bf]{subcaption}

\usepackage{tikz,pgfplots}
\usetikzlibrary{calc}
\usetikzlibrary{intersections}
\usetikzlibrary{arrows,shapes}
\usetikzlibrary{positioning}

\newtheorem{lemma}{Lemma}

\theoremstyle{definition}

\allowdisplaybreaks

\begin{document}
\title{
Maximizing Spectral and Energy Efficiency in Multi-user MIMO OFDM Systems with \\RIS and Hardware Impairment
}
\author{Mohammad Soleymani, \emph{Member, IEEE},  
Ignacio Santamaria, \emph{Senior Member, IEEE}, \\
Aydin Sezgin, \emph{Senior Member, IEEE}, 
and
Eduard Jorswieck, \emph{Fellow, IEEE}
 \\ \thanks{ 
A preliminary version of a portion of this work was presented in
\emph{Proc. of IEEE CAMSAP 2023} \cite{soleymani2023maximization}.

Mohammad Soleymani is with the Signal and System Theory Group, Universit\"at Paderborn, Germany, (email: \protect\url{mohammad.soleymani@uni-paderborn.de}).  

Ignacio Santamaria is with the Department of Communications Engineering, Universidad de Cantabria (email: \protect\url{i.santamaria@unican.es}).

Aydin Sezgin is with the  Ruhr University Bochum, Germany (email: \protect\url{aydin.sezgin@rub.de}).

Eduard Jorswieck is with the Institute for Communications Technology, Technische Universit\"at Braunschweig, 38106 Braunschweig, Germany
(e-mail: \protect\url{jorswieck@ifn.ing.tu-bs.de})

The work of Ignacio Santamaria was funded by MCIN/ AEI /10.13039/501100011033, under Grant PID2022-137099NB-C43 (MADDIE). 
This work of Aydin Sezgin is funded by the German Federal Ministry of Education and
Research (BMBF) in the course of the 6GEM Research Hub under Grant
16KISK037. The work of Eduard Jorswieck was supported by the Federal Ministry of Education and Research (BMBF, Germany) through the Program of ``Souver\"an. Digital. Vernetzt.'' joint Project 6G-RIC, under Grant 16KISK031. 
}}
\maketitle
\begin{abstract} An emerging technology to enhance the spectral efficiency (SE) and energy efficiency (EE) of wireless communication systems is reconfigurable intelligent surface (RIS), which is shown to be very powerful in single-carrier systems. However,  in multi-user orthogonal frequency division multiplexing (OFDM) systems, RIS may not be as promising as in single-carrier systems since an independent optimization of RIS elements  at each sub-carrier is impossible in multi-carrier systems. Thus, this paper investigates the performance of various RIS technologies like regular (reflective and passive), simultaneously transmit and reflect (STAR), and multi-sector beyond diagonal (BD) RIS in multi-user multiple-input multiple-output (MIMO) OFDM broadcast channels (BC). This requires to formulate and solve a joint MIMO precoding and RIS optimization problem. The obtained solution reveals that RIS can significantly improve the system performance even when the number of RIS elements  is relatively low.  Moreover, we develop resource allocation schemes for STAR-RIS and multi-sector BD-RIS in MIMO OFDM BCs, and show that these RIS technologies can outperform a regular RIS, especially when the regular RIS cannot assist the communications for all the users.
\end{abstract} 
\section{Introduction}
Smart radio environments can be realized by employing modern intelligent metasurface technologies \cite{renzo2019smart}. Indeed, wireless channels in systems aided by reconfigurable intelligent surfaces (RIS) are not determined by nature only and can be optimized \cite{wu2019towards}. This can yield additional degrees of freedom (DoF) in designing systems, which  can be utilized to substantially improve the performance of wireless systems, especially when the system is single carrier \cite{wu2021intelligent, di2020smart}. However, in multi-carrier systems, the sub-carriers undergo different channel states. Thus,  the relative improvement per carrier might be lower since RIS elements cannot be optimized independently at each sub-carrier. Hence, it should be investigated how well RIS can perform in multi-carrier systems in which one is unable to optimize each carrier independently of other carriers.  
In this paper, we address this issue and propose a framework to enhance the spectral efficiency (SE) and energy efficiency (EE) of multi-user multiple-input, multiple-output (MIMO) orthogonal frequency division multiplexing (OFDM) systems by jointly optimizing the power/transmit covariance matrices and the RIS elements. 
Note that SE and EE improvements are always among the main concerns of modern wireless communication systems, where a goal in 6G is to achieve SE  10 times and EE 100 times higher  than 5G systems \cite{wang2023road, gong2022holographic}.

This work shows that the RIS benefits in multi-user MIMO OFDM systems decrease with the number of sub-carrier/sub-bands, but the benefits are still significant even when the number of RIS elements per user/sub-carrier is very low. 
\subsection{Literature review}
Note that there are various technologies for intelligent metasurfaces. The simplest architecture is the passive reflective RIS, which is referred to as a regular RIS in this paper. It has been shown that RIS can be a powerful technology to improve the SE and EE of several single-carrier systems \cite{zappone2020optimal, huang2019reconfigurable, wu2019intelligent, soleymani2022rate, soleymani2023rate, pan2020multicell, soleymani2022improper, soleymani2022noma, soleymani2023energy, zhang2020intelligent, jiang2022interference, santamaria2023icassp, weinberger2023ris, esmaeili2022fairness, weinberger2022synergistic, soleymani2023optimization, santamaria2023snr}. For example, the papers in \cite{huang2019reconfigurable, wu2019intelligent} illustrated that RIS can improve the performance of a single-cell multiple-input single-output (MISO) broadcast channel (BC). In  \cite{ soleymani2022rate, soleymani2022improper}, the superiority of RIS was examined in multi-cell MIMO BCs by considering different performance metrics such as the minimum and sum rate, global EE and minimum EE of users when transceivers suffer from hardware impairment (HWI). The authors in \cite{zhang2020intelligent} showed that RIS can increase the weighted sum rate of the secondary users in a MIMO cognitive radio system. In \cite{  soleymani2023rate}, it was demonstrated that RIS can enlarge the achievable rate region of a single-cell MIMO BC. In \cite{jiang2022interference, santamaria2023icassp}, RIS was employed as a tool to reduce the interference leakage of the $K$-user interference channels. The paper in \cite{weinberger2023ris} exhibited that RIS can enhance the resilience of cell-free MIMO systems. 

A drawback of a regular RIS is that it cannot provide an omni-directional coverage. In other words, the transmitter and receiver have to be in the reflection space of a regular RIS so that the regular RIS can bring any benefit to the system. Thus, a regular RIS supports only a half-space coverage. To address this issue, another technology for intelligent metasurfaces is proposed that can allow transmission and reflection at the same time, which in known as simultaneously transmit and reflect (STAR) RIS \cite{liu2021star}. Note that STAR-RIS is also known as intelligent omni-surfaces since it can provide an omni-directional coverage \cite{liu2021star, zhang2022intelligent}. The superiority of STAR-RIS over regular RIS has been investigated in \cite{9774942, mu2021simultaneously, xu2021star, soleymani2023energy, soleymani2023noma, soleymani2023spectral} by taking into account different scenarios. For instance, in \cite{9774942, mu2021simultaneously, xu2021star}, it was shown that STAR-RIS can improve the performance of a two-user BC with single-antenna receivers. In \cite{soleymani2023energy}, we proposed schemes to increase the EE of a single-cell MIMO STAR-RIS-aided BC. In \cite{soleymani2023noma}, it was demonstrated that STAR-RIS can increase the minimum rate of users in a multi-cell BC with I/Q imbalance (IQI). Moreover, in \cite{soleymani2023spectral}, it was shown that STAR-RIS can enhance the performance of MISO ultra-reliable low-latency communications (URLLC) BCs. 

The concept of STAR-RIS can further be generalized to multi-sector beyond diagonal (BD) RIS in which the coverage area is divided into multiple sectors (possibly more than two) \cite{li2023reconfigurable, li2023beyond}. In this case, the multi-sector BD-RIS can receive  signal in a sector and partially reflect it in all the sectors, which enables a full-space coverage similar to STAR-RIS. In this sense, multi-sector BD-RIS can be also categorized as omni-directional surfaces. Note that in BD-RIS, the matrix for the RIS coefficients can be  non-diagonal since the BD-RIS elements can be connected via a circuit design. In this paper, we consider the single-connected multi-sector BD-RIS in which the RIS elements in each sector are not connected to the other RIS elements in the sector. It should be emphasized that the main advantage of  multi-sector BD-RIS over STAR-RIS is that the multi-sector BD-RIS can provide more directional beams in each sector, which can enhance the channel gain \cite{ li2023beyond}. Additionally, the channel of a user in a sector can be optimized independently of the channels of the users in the other sectors. This simplifies optimization of the channels of the users and enhance the performance of RIS. 

Yet, one of the main factors limiting the performance of wireless communication systems is HWI. In practice, devices never perform ideally, and if we do not take the device imperfections into consideration when designing a system, the system performance can drop significantly \cite{soleymani2019improper, soleymani2020improper, javed2019multiple, soleymani2020rate}. A common source of HWI is I/Q imbalance (IQI), which happens because of an imbalance in the in-phase and quadrature signals \cite{soleymani2020improper, javed2019multiple}. IQI is modeled as a widely linear transformation (WLT), which makes the output signal improper when the input signal is proper \cite{soleymani2020improper, javed2019multiple}. To compensate for IQI, one can resort to  improper Gaussian signaling (IGS), which is also an effective interference-management tool, especially in single-carrier systems \cite{cadambe2010interference, soleymani2019improper, soleymani2019robust}. 
However, the benefits of IGS as an interference-management technique may disappear in multi-carrier systems as the number of resources per users increases, which in turn may help to easier manage interference by a simpler transmission strategy \cite{soleymani2018improper}.

\begin{table}
\centering
\footnotesize
\caption{Summary of the most related papers on OFDM RIS-aided systems.}\label{table-1}
\begin{tabular}{|l|c|c|c|c|c|c|c|c|c|}
	\hline
 & This paper& \cite{li2023piecewise, yue2023ris, wu2022intelligent, kompostiotis2023received, li2021intelligent, dong2022sparse, pradhan2020reconfigurable, ji2023joint, jiang2023bivariate, abbas2023unit} &\cite{peng2022optimization, wang2023joint} & \cite{ginige2021untrained}
\\
\hline
MIMO&$\surd$&&$\surd$&
\\
\hline
EE metrics&$\surd$&&&
\\
\hline
Fairness rate&$\surd$&&&
\\
\hline
Multi-user&$\surd$&$\surd$&&
\\
\hline
HWI&$\surd$&&&$\surd$
\\
\hline
IGS&$\surd$&&&
\\
\hline
STAR-RIS&$\surd$&&&
\\
\hline
Multi-sector BD-RIS&$\surd$&&&
\\
\hline
		\end{tabular}
\normalsize
\end{table} 
Now, single-carrier systems are very efficient in frequency-flat channels. However, this might not necessarily be the case when the channels are frequency selective since a demanding process is needed for equalizing the channels in single-carrier systems. To easier cope with frequency selectivity of wireless channels,  
multi-carrier techniques such as OFDM can be employed, which are able to more efficiently utilize the spectrum. Multi-carrier systems divide the frequency band into several sub-bands, which permits a more intelligent power allocation, based on the channel response on each frequency sub-band. However, RIS cannot exploit this feature since it is impossible to independently optimize RIS at each sub-band \cite{li2021intelligent}. In a more realistic scenario, the RIS elements remain approximately constant across all frequency subbands, which is highly suboptimal when the number of subbands grows. Thus, one might expect that the benefits of RIS may disappear when the bandwidth is large, and consequently, the number of subbands is high. 

Although limited, yet a few studies exist on RIS-aided multi-user OFDM systems  \cite{zhang2022reconfigurable, lin2020adaptive, he2021low, bjornson2021optimizing, zheng2019intelligent,  yang2020risofdm, he2022low, an2022scalable, xu2022sparse, huang2023novel, yang2021channel, he2021Cooperative, yan2022frequency, li2023piecewise, yue2023ris, wu2022intelligent, kompostiotis2023received, li2021intelligent, dong2022sparse, pradhan2020reconfigurable, ji2023joint, jiang2023bivariate, abbas2023unit, peng2022optimization, wang2023joint, ginige2021untrained}. 
Many studies on RIS-aided OFDM systems considered a single-user system \cite{zhang2022reconfigurable, lin2020adaptive, he2021low, bjornson2021optimizing, zheng2019intelligent,  yang2020risofdm, he2022low, an2022scalable, xu2022sparse, huang2023novel, yang2021channel, he2021Cooperative, yan2022frequency,
peng2022optimization, wang2023joint,
ginige2021untrained}. 
For instance, in \cite{lin2020adaptive}, the authors proposed a joint channel estimation and passive beamforming scheme
for a  RIS-aided OFDM system and showed that RIS can substantially improve the SE of the system.
 A low-cost passive beam forming was proposed in \cite{he2021low} for point-to-point SISO RIS-aided OFDM systems, and it was shown that RIS can increase the achievable rate of the system. 
The authors in \cite{zheng2019intelligent} considered a point-to-point single-input single-output (SISO) RIS-aided OFDM system and proposed schemes to estimate channels as well as to optimize RIS elements. In \cite{yang2020risofdm}, the authors developed a practical transmission protocol for point-to-point SISO RIS-aided OFDM systems. 
The authors in \cite{peng2022optimization} proposed schemes to optimize a MIMO OFDM simultaneous wireless information and power transfer
(SWIPT) RIS-aided system with a non-linear energy harvesting model and showed that RIS can enhance the system performance. 
In \cite{wang2023joint}, the authors considered a point-to-point MIMO RIS-aided OFDM system and showed that RIS can improve the SE of the system.

The performance of RIS in a multi-user OFDM system has been studied in \cite{li2023piecewise, yue2023ris, wu2022intelligent, kompostiotis2023received, li2021intelligent, dong2022sparse, pradhan2020reconfigurable, ji2023joint, jiang2023bivariate}. In \cite{li2021intelligent, wu2022intelligent}, it was shown that RIS can increase the sum-rate of multi-user multiple-input single-output (MISO) RIS-aided OFDM systems. In \cite{kompostiotis2023received}, the authors studied a multi-user SISO RIS-aided OFDM BC and optimized the RIS elements to maximize the total power received by users for a given transmit power. The authors in \cite{yue2023ris} considered an uplink of a multi-user OFDM system, and showed that RIS can decrease the bit error rate of the system. 

\subsection{Motivation}
We provide a brief summary for  the papers on RIS in OFDM systems in Table \ref{table-1}, based on the system model, the considered performance metric,  assumptions regarding the devices, and the considered RIS technology.  Even though the studies in \cite{zhang2022reconfigurable, lin2020adaptive, he2021low, bjornson2021optimizing, zheng2019intelligent,  yang2020risofdm, he2022low, an2022scalable, xu2022sparse, huang2023novel, yang2021channel, he2021Cooperative, yan2022frequency, li2023piecewise, yue2023ris, wu2022intelligent, kompostiotis2023received, li2021intelligent, dong2022sparse, pradhan2020reconfigurable, ji2023joint, jiang2023bivariate, peng2022optimization, wang2023joint} have provided a valuable insight on the performance of RIS in OFDM systems, there are still some open problems in RIS-aided OFDM systems, which are the focus of this paper. For instance, the performance of RIS has not been studied in multi-user MIMO OFDM systems. Most of the modern wireless communication systems such as 4G and 5G employ MIMO and OFDM, which suggests developing resource management techniques for RIS-aided MIMO OFDM systems.
 Additionally,  there is no work on RIS-aided OFDM systems that considers the EE metrics,  even in single-antenna, single-user OFDM systems. As indicated before, a main goal of 6G is to increase EE by an order of approximately 100 times compared to 5G systems, which makes energy-efficient schemes vital for 6G. Furthermore, there is only one paper (i.e., \cite{ginige2021untrained}) on RIS-aided OFDM systems, which considered a single-user user system with additive hardware distortion noise. Indeed, IQI has not been studied in RIS-aided OFDM systems, while in practice, devices are never ideal and may suffer from IQI and/or other sources of imperfections. Moreover, there is no work on RIS-aided OFDM systems with more recent RIS technologies such as STAR-RIS or  multi-sector BD-RIS. 

As a summary, there are still some open research lines in MU-MIMO RIS-aided OFDM systems, which require further investigations. These include developing HWI-aware designs, energy-efficient techniques and fairness-aware schemes. Moreover, there are some new technologies and/or concepts for RIS that have not been studied in MU-MIMO OFDM systems.

\subsection{Contribution}
In this paper, we maximize the SE and the EE of multi-user MIMO RIS-aided OFDM systems by considering various technologies for RIS like regular, STAR and multi-sector BD-RIS. To the best of our knowledge, this is the first work on RIS-aided OFDM systems that considers EE metrics and proposes IQI-aware schemes for RIS-aided OFDM systems. In the conference version of this work \cite{soleymani2023maximization}, we proposed optimization schemes for maximizing the minimum rate of users in a single-cell MIMO RIS-aided BC with perfect devices.  This paper extends the results in \cite{soleymani2023maximization} to multi-cell MIMO RIS-aided BC with IQI, and considers other technologies for RIS, including STAR-RIS and multi-sector BD-RIS. 

The proposed algorithms in this work can be applied to a wide range of optimization problems in which the objective function and/or constraints are linear functions of the rates and/or EE of users. Such optimization problems include many practical objective functions such as minimum and/or sum rate, global EE, and minimum EE. Furthermore, we can apply the algorithms to any multi-user MIMO RIS-aided OFDM system with IQI at transceivers. In this work, we consider a multi-cell MIMO RIS-aided OFDM BC as an example of such a system and show that RIS (either regular, STAR, or multi-sector) can substantially increase the SE and EE even with a relatively low number of RIS components. In more detail, our numerical results show that optimizing the RIS elements can provide a high gain in OFDM systems even when we cannot independently optimize the RIS elements at each subband. These results show that RIS can be a promising technology to enhance the performance of OFDM systems.

We, additionally, propose IQI-aware schemes for MU MIMO RIS-aided OFDM systems. To this end, we employ IGS since IQI makes the signals improper if proper signaling is employed, which reduces the achievable rate \cite{soleymani2020improper}. We can also employ IGS as an interference-management technique, especially in single-carrier systems. However, as shown in \cite{soleymani2018improper}, the IGS benefits  as an interference-management tool vanish when the number of subbands increases. Indeed, the higher the number of subbands is, the more resources per user are available, which reduces the interference level. Hence, the IGS gains in this scenario are mainly because of the ability to compensate for IQI. Note that IQI can significantly decrease the system performance, especially when we neglect it in designing signaling schemes.

Finally, we develop optimization schemes for STAR-RIS as well as  multi-sector BD-RIS in multi-user MIMO OFDM BCs. We show that STAR-RIS even with random coefficients can outperform a regular RIS with optimized elements. Additionally, it is shown that optimizing STAR-RIS elements can result in substantial improvements, and STAR-RIS can highly outperform a regular RIS when all the users are not in the coverage area of the regular RIS. Moreover, our results show that multi-sector BD-RIS can significantly outperform STAR-RIS. We consider different schemes for operating STAR-RIS and multi-sector BD-RIS. Firstly, we assume that all the multi-sector BD-RIS elements can operate in all sectors simultaneously, which we refer to as the energy splitting (ES) scheme. Secondly, we assume that each multi-sector BD-RIS element can operate in only one sector at a time, which is called the mode switching (MS) scheme. We show that the performance of the MS scheme is comparable with ES, while it has much lower computational and implementation complexities.

\section{System model}\label{sec-ii}
We consider a multi-cell BC consisting of $L$ multiple-antenna BS with $N_{\mathsf{B}}$ antennas each. In each cell, there are $K$ users with $N_{\mathsf{U}}$ antennas each. Moreover, there are $N$ RISs (either multi-sector, STAR or regular) with $N_{\mathsf{R}}$ elements to assists the BS, and there is at least one RIS in each cell. We assume that each BS transmits a linear superposition of multicarrier OFDM signals with $N_i$ sub-carriers.

\subsection{RIS model}\label{seciia}
This paper studies different RIS technologies such as regular, STAR and multi-sector in a multi-user MIMO OFDM system. Below, we first describe the concept of STAR-RIS and multi-sector BD-RIS briefly. Then we state the effective channels in MIMO OFDM systems, aided by these RIS technologies. Moreover, we explain different operational modes for a multi-sector BD-RIS and/or STAR-RIS. 

\subsubsection{Concept of multi-sector BD-RIS}
\begin{figure}%[t!]
    \centering
    \begin{subfigure}{0.4\textwidth}
        \centering
\includegraphics[width=\textwidth]{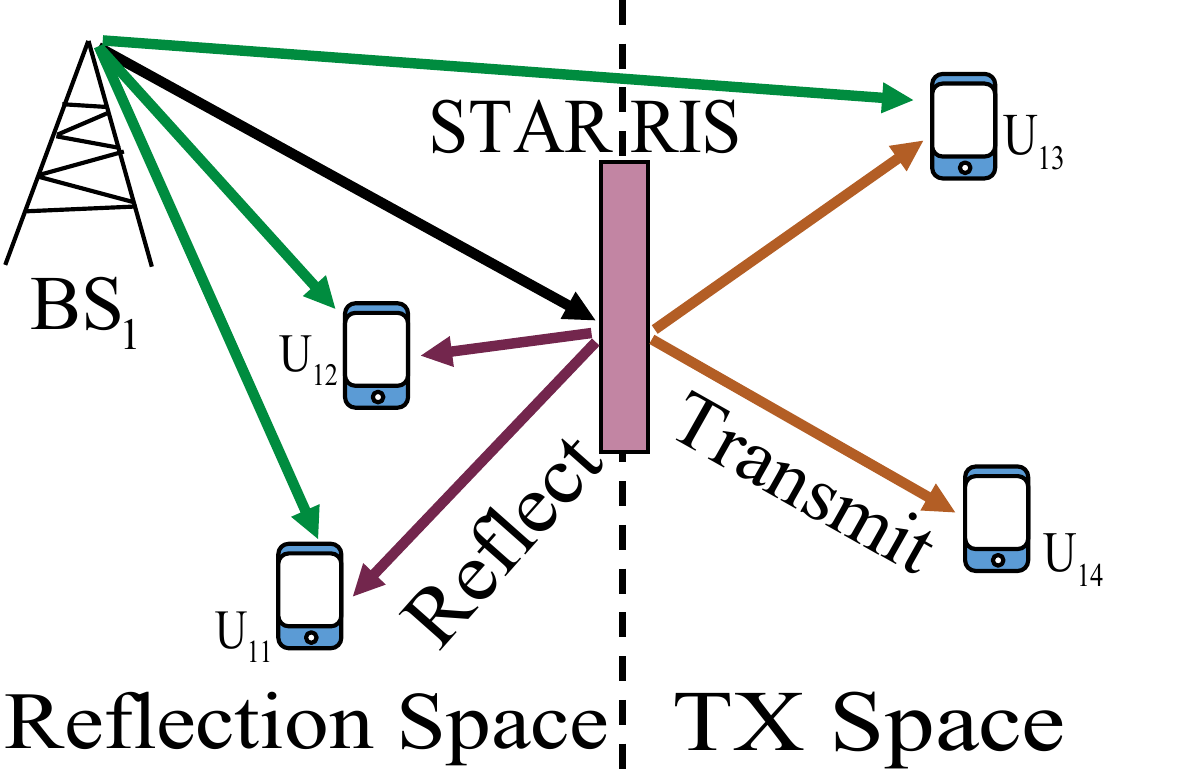}
     \caption{STAR-RIS ($N_s=2$).}
    \end{subfigure}
\\
	\begin{subfigure}{0.4\textwidth}
        \centering
           \includegraphics[width=.7\textwidth]{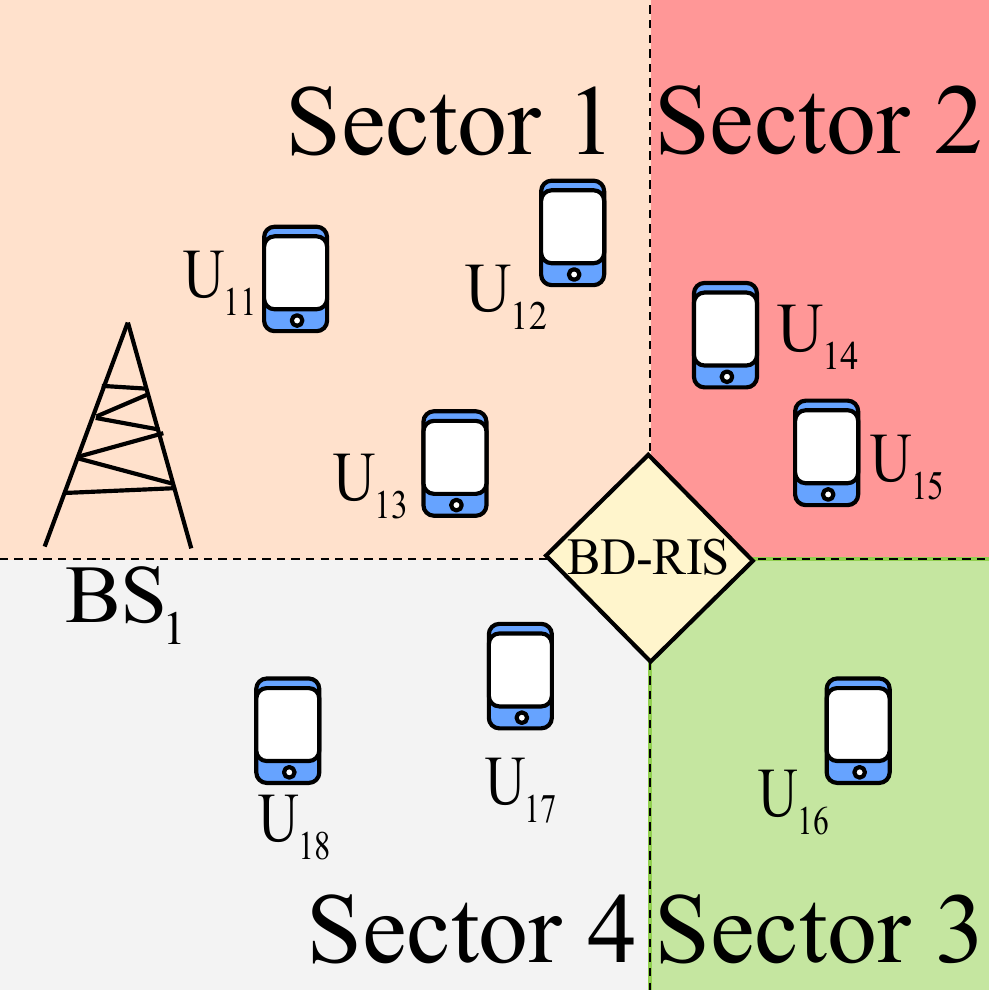}
        \caption{Four-sector BD-RIS ($N_s=2$).}
    \end{subfigure}
\caption{A broadcast channel aided by a multi-sector BD-RIS.}
	\label{Fig-star}
\end{figure}

Multi-sector BD-RIS is  a generalization of STAR-RIS and can be categorized as intelligent omni-surfaces, which can provide a full $360^\circ$ coverage. In a STAR-RIS, each element can  transmit and reflect simultaneously, which realizes an omni-directional coverage \cite{zhang2022intelligent}. Thus, there are two spaces for each STAR-RIS, which are referred to as the reflection space and the transmission space (see Fig. \ref{Fig-star}a). 
The concept/technology of STAR-RIS can be extended to multi-sector BD-RIS in which each RIS can have multiple sectors, each covering a set of users and/or a sub-space, as shown in Fig. \ref{Fig-star}b. 
In the most general case, a multi-sector BD-RIS can have $N_s$ sectors, and when $N_s=2$, the multi-sector BD-RIS is equivalent to a STAR-RIS. 
Thus, in a multi-sector-BD-RIS-aided system, each user belongs to a region/space, covered by a sector of  the multi-sector BD-RIS. For instance, in Fig. \ref{Fig-star}b, users u$_{11}$, u$_{12}$ and u$_{13}$ are covered by the first sector of the multi-sector BD-RIS, while users u$_{17}$ and u$_{18}$ are covered by the fourth sector of the multi-sector BD-RIS.

\subsubsection{Effective channels in frequency domain}
We consider the channels in frequency domain. To this end, we employ the channel model in \cite[Eq. (12)]{li2021intelligent}. Thus, the channel between the $j$-th BS and the $k$-th user associated to the $l$-th BS, denoted by u$_{lk}$, at subband $i$ is 
\begin{equation}\label{ch-equ}
\mathbf{H}_{lk,j,i}\!\left(\{\bm{\Phi}\}\right)\!=\!
{\sum_{n=1}^N\!\mathbf{G}_{lk,n,i}\bm{\Phi}_n^{n_s}\mathbf{G}_{nj,i}}
\!+\!
{\mathbf{F}_{lk,j,i}}
\end{equation}
where $\mathbf{G}_{lk,n,i}$ %\in\mathbb{C}^{N_{\mathsf{U}}\times N_{\mathsf{R}}}$ 
is the channel between u$_{lk}$ and the $n$-th RIS in subband $i$, 
$\mathbf{G}_{nj,i}$ %\in\mathbb{C}^{N_{\mathsf{R}}\times N_{\mathsf{B}}}$ 
is the channel between the $n$-th multi-sector BD-RIS and the $j$-th BS in subband $i$, $\mathbf{F}_{lk,j,i}$ %\in\mathbb{C}^{N_{\mathsf{U}}\times N_{\mathsf{B}}}$ 
is the direct link between u$_{lk}$ and the $j$-th BS in subband $i$. Moreover, the matrix $\bm{\Phi}_n^{n_s}$ contains the coefficients for the $n_s$-th sector of the multi-sector BD-RIS in which u$_{lk}$ is located. The matrices $\bm{\Phi}_n^{n_s}$s for $n_s=1,\cdots,N_s$ are diagonal as 
\begin{align}
\bm{\Phi}^{n_s}_n&=\text{diag}\left(\phi_{n1}^{n_s}, \phi_{n2}^{n_s},\cdots,\phi_{n{N_{\mathsf{R}}}}^{n_s}\right)\hspace{1cm}\forall n_s,m.
\end{align}
Note that the multi-sector BD-RIS, which is considered in this paper, is referred to as  the cell-wise single-connected multi-sector BD-RIS in \cite{li2023beyond}. 
To operate in a passive mode, the amplitudes of the coefficients for each multi-sector BD-RIS element should satisfy 
\begin{equation}\label{sr-1}
\sum_{n_s=1}^{N_s}|\phi_{nm}^{n_s}|^2\leq 1, \hspace{1cm}\forall n,m,
\end{equation}
which is a convex constraint. We represent the feasibility set, corresponding to this constraint as
\begin{equation}\label{model-2}
\mathcal{T}_{U}=\left\{\phi_{nm}^{n_s}:\sum_{n_s=1}^{N_s}|\phi_{nm}^{n_s}|^2\leq 1 \,\,\,\forall n,m\right\}.
\end{equation}
Assuming passive and lossless operation for each multi-sector BD-RIS, we have the following constraint 
\cite[Eq. (3)]{li2023synergizing}
\begin{equation}\label{sr-2}
\sum_{n_s=1}^{N_s}|\phi_{nm}^{n_s}|^2= 1, \hspace{1cm}\forall n,m.
\end{equation}
 We represent the feasibility set for the constraint \eqref{sr-2} as
\begin{equation}\label{model-1}
\mathcal{T}_{I}=\left\{\phi_{nm}^{n_s}:\sum_{n_s=1}^{N_s}|\phi_{nm}^{n_s}|^2= 1 \,\,\,\forall n,m\right\}.
\end{equation}
In these two models, it is  assumed that each multi-sector BD-RIS operates in a passive mode, which yields a constraint only on the amplitude of the multi-sector  BD-RIS elements. However,
there might be some additional constraints on the phases of the coefficients for each multi-sector BD-RIS component in practice. Unfortunately, multi-sector BD-RIS has not been implemented yet, and we have to consider only $\mathcal{T}_{U}$ and $\mathcal{T}_{I}$. Nevertheless,  according to the model in \cite{9774942}, the phases of the reflection and transmission coefficients for each STAR-RIS element are highly dependent as $\angle\phi_{nm}^1=\angle\phi_{nm}^2\pm\frac{\pi}{2}$, which results in 
\begin{multline}\label{model-3}
\mathcal{T}_{SN}=\left\{\phi_{nm}^1,\phi_{nm}^2:|\phi_{nm}^{1}|^2+|\phi_{nm}^{2}|^2= 1,\right.
\\
\left. \mathfrak{R}\left\{\phi_{nm}^{1^*}\phi_{nm}^{2}\right\}= 0\,\,\,\forall n,m\right\}.
\end{multline}
According to \cite[Lemma 1]{soleymani2023spectral}, the constraints for each STAR-RIS component in the set $\mathcal{T}_{SN}$ can be rewritten as
\begin{align}\label{st-eq-}
|\phi_{nm}^{1}+\phi_{nm}^{2}|^2&\leq 1,\\
\label{st-eq-1}
|\phi_{nm}^{1}-\phi_{nm}^{2}|^2&\leq 1,\\
|\phi_{nm}^{1}|^2+|\phi_{nm}^{2}|^2&= 1.
\label{st-eq-2}
\end{align}
Indeed, $\mathcal{T}_{SN}$ is only valid for STAR-RIS ($N_s=2$) and is more stringent than the other feasibility sets since it has additional constraints on the phases of the STAR-RIS components. 
Moreover, it can be easily verified that $\mathcal{T}_{SN}\subset\mathcal{T}_{I}\subset\mathcal{T}_{U}$. Hereafter, we represent the feasibility set of RIS coefficients by $\mathcal{T}$, unless we explicitly refer to one  of these feasibility sets.

Note that according to the model in \cite{li2021intelligent}, the reflecting coefficients cannot be independently optimized at each subband. Thus, we assume that the coefficients remain constant across all the frequency subbands. 
Additionally, it should be noted that as can be verified through \eqref{ch-equ}, the channels are linear functions of $\{\bm{\Phi}\}=\{\bm{\Phi}_n^{n_s}\}_{\forall n_s,n}$. However, for notational simplicity, we do not state this explicitly hereafter.

\subsubsection{Operational modes for multi-sector BD-RIS} 
There are different possibilities to operate a multi-sector BD-RIS. For instance, all the multi-sector BD-RIS components can actively operate in all sectors, which is referred to as the ES scheme. This scheme is the most general case to operate a multi-sector BD-RIS, and is expected to outperform the other operational schemes. However, the ES scheme has also higher computational and implementation complexities. In this mode, there are $N_s$ complex-valued optimization parameters per multi-sector-BD-RIS element, which can provide more design flexibility at the cost of higher computational complexities. Additionally, operating each  component of the multi-sector BD-RIS in all sectors may require  more advanced circuit designs. As indicated, the coefficients corresponding to each sector might be also highly dependent not only through the amplitudes, but also through the phases, which may make ES schemes inefficient in practice. 

To cope with the challenges of ES schemes, we can consider different operational modes for multi-sector BD-RIS with lower complexities. For instance, each component of the multi-sector BD-RIS can operate only in a sector at a time. To realize such schemes, one possibility can be to divide the components of the multi-sector BD-RIS into $N_s$ groups. In the $n_s$-th group, all the  components of the multi-sector BD-RIS operate only in the $n_s$-th sector, which is called the MS scheme. To further clarify MS, we provide an example for $N_s=2$, which is equivalent to a STAR-RIS. In this case, the components are divided into two groups.   In the first group, all the STAR-RIS components operate only in the reflection mode, while the STAR-RIS elements in the other group operate only in the transmission mode. Another possibility could be to   operate all the elements of a multi-sector BD-RIS in one sector in a time slot, and switch the operating sector in next time slots in a round robin to cover all the sectors, which we call the time switching (TS) scheme. In a STAR-RIS, it means that all the STAR-RIS elements operate in a reflection mode in a time slot, while they all operate in a transmission mode in the next time slot. Of course, there can be also hybrid schemes, which are a combination of the ES, MS and TS schemes, but we do not consider such schemes in this work. 

The ES and MS schemes can provide a full coverage at a time, while the TS scheme covers only a subspace and is unable to provide a full coverage in a single time slot. As a result, we consider only the ES and MS schemes, but our proposed schemes can be easily applied to TS and/or hybrid schemes.

\subsection{I/Q imbalance model}\label{iqi-sec}
We consider the IQI model in \cite{soleymani2020improper}. Note, that the model in \cite{soleymani2020improper} considers single-carrier systems, however we can be extend it to multi-carrier systems. If the system at hand is wide-band, it is expected that the IQI parameters differ at each subband, based on \cite[Table I]{sha2021channel}. However, in narrow-band systems, the IQI parameters can be the same in all subbands. In this paper, we investigate the most general case in which each subband can have different IQI parameters. We briefly restate the model in \cite{soleymani2020improper} for the sake of completeness below.
 
When IQI occurs at a device, the output signal is a WLT of the input signal.  
Hence, if we denote the input signal of a MIMO OFDM system at subband $i$ as ${\bf x}_i$, the actual transmitted signal is 
\begin{equation}\label{iqi-tx}
{\bf x}_{t,i}=\mathbf{\Gamma}_{t1,i}{\bf x}_i+\mathbf{\Gamma}_{t2,i}{\bf x}^*_i,
\end{equation}
where ${\bf x}^*_i$ is the conjugate of ${\bf x}_i$, and the coefficients $\mathbf{\Gamma}_{t1,i}$ and $\mathbf{\Gamma}_{t2,i}$ are, respectively, given by
\begin{align}
\mathbf{\Gamma}_{t1,i}&=\frac{\mathbf{I}+\mathbf{A}_{T,i}e^{j \bm{\psi}_{T,i}}}{2},&
\mathbf{\Gamma}_{t2,i}&={\bf I}-\mathbf{\Gamma}_{t1,i}^*,%=\frac{\mathbf{I}-\mathbf{A}_Te^{-j \bm{\phi}_T}}{2},
\end{align}
where $\mathbf{A}_{T,i}=a_{t,i}{\bf I}_{N_{\mathsf{U}}}$ and $\bm{\psi}_{T,i}=\psi_{t,i}{\bf I}_{N_{\mathsf{U}}}$ are, respectively, the coefficients corresponding to the amplitude and phase mismatches at the transmitter side. Note that $a_{t,i}$ and $\psi_{t,i}$ are real-valued scalar parameters, and the device is ideal if $a_{t,i}=1$ and $\psi_{t,i}=0$.

At the receiver side, we have a similar model in which the output signal is a WLT of the received signal as
\begin{equation}\label{iqi-rx}
{\bf y}_{i}=\mathbf{\Gamma}_{r1,i}{\bf y}_{r,i}+\mathbf{\Gamma}_{r2,i}{\bf y}^*_{r,i},
\end{equation}
where ${\bf y}_{r,i}={\bf H}_{i}{\bf x}_{t,i}+{\bf n}_{i}$ is the received signal, where ${\bf H}_{i}$ and ${\bf n}_{i}$ are, respectively, the MIMO channel, and  additive noise at the $i$-th subband. 
Moreover, the coefficients $\mathbf{\Gamma}_{r1,i}$ and $\mathbf{\Gamma}_{r2,i}$ are 
\begin{align}
\mathbf{\Gamma}_{r1,i}&=\frac{\mathbf{I}+\mathbf{A}_{R,i}e^{j \bm{\phi}_{R,i}}}{2},&
\mathbf{\Gamma}_{r2,i}&={\bf I}-\mathbf{\Gamma}_{r1,i}^*,%=\frac{\mathbf{I}-\mathbf{A}_Te^{-j \bm{\phi}_T}}{2},
\end{align}
respectively, where $\mathbf{A}_{R,i}=a_{r,i}{\bf I}_{N_{\mathsf{B}}}$ and $\bm{\phi}_{R,i}=\phi_{r,i}{\bf I}_{N_{\mathsf{B}}}$ are, respective, the coefficients corresponding to the amplitude and phase mismatches at the receiver side. Similarly, $a_{r,i}$ and $\phi_{r,i}$ are real-valued scalar parameters, and the receiver is perfect if $a_{r,i}=1$ and $\phi_{r,i}=0$.
Finally, the signal at the output of the receiver is 
\begin{multline}\label{8}
{\bf y}_{i}=
\left(\mathbf{\Gamma}_{r1,i}{\bf H}_{i}\mathbf{\Gamma}_{t1,i}
+\mathbf{\Gamma}_{r2,i}{\bf H}_{i}^*\mathbf{\Gamma}_{t2,i}^*
\right){\bf x}_{t,i}
\\+
\left(\mathbf{\Gamma}_{r1,i}{\bf H}_{i}\mathbf{\Gamma}_{t2,i}
+\mathbf{\Gamma}_{r2,i}{\bf H}_{i}^*\mathbf{\Gamma}_{t1,i}^*
\right){\bf x}_{t,i}^*
\\+
\mathbf{\Gamma}_{r1,i}{\bf n}_{i}+\mathbf{\Gamma}_{r2,i}{\bf n}^*_{i}.
\end{multline}
In the following lemma, we restate \eqref{8} in real vector-valued variables by employing the real-decomposition method.
\begin{lemma}[\!\!\cite{soleymani2020improper}]\label{lem-iqi}
Equation \eqref{8} can be rewritten in a real domain as
\begin{equation*}
\underline{\mathbf{y}}_{i}=\underline{\mathbf{H}}_{i}\underline{\mathbf{x}}_{i}+\underline{\mathbf{n}}_{i},
\end{equation*}
where $\underline{\mathbf{y}}_i$, $\underline{\mathbf{x}}_i$, and $\underline{\mathbf{n}}_i$ are, respectively, the real decomposition of ${\bf y}_i$, ${\bf x}_i$, and $\mathbf{\Gamma}_{r1,i}{\bf n}_{i}\!+\!\mathbf{\Gamma}_{r2,i}{\bf n}^*_{i}$. 
Note that the effective noise at the output of the receiver is $\underline{\mathbf{n}}_i$, which is zero-mean Gaussian with covariance matrix
$\underline{\mathbf{C}}_i= \sigma^2\underline{\mathbf{\Gamma}}_i
\underline{\mathbf{\Gamma}}^T_i,$
where $\sigma^2$ is the noise variance at each received antenna, and $\underline{\mathbf{\Gamma}}_i$ can be obtained as in \cite[Eq. (13)]{soleymani2022improper}.
Moreover, the equivalent channel $\underline{{\bf H}}_i$ is given by \cite[Eq. (11)]{soleymani2022improper}. \end{lemma}

\subsection{Signal model}
BS $l$ intends to transmit 
\begin{equation}
\mathbf{x}_{l,i}=\sum_{k=1}^K\mathbf{x}_{lk,i},%\in\mathbb{C}^{N_{\mathsf{B}}\times 1},
\end{equation}
at subband $i$,
where $\mathbf{x}_{lk,i}$ %\in\mathbb{C}^{N_{\mathsf{B}}\times 1}$ 
is the transmit  signal of BS $l$ intended for u$_{lk}$ at subband $i$.  As described in Section \ref{iqi-sec}, the signal that BS $l$ transmits is a WLT of  $\mathbf{x}_{l,i}$, based on \eqref{iqi-tx}. 
Employing Lemma \ref{lem-iqi}, the received signal received by u$_{lk}$ in subband $i$ is
\begin{align}
\nonumber
\underline{\mathbf{y}}_{lk,i}&=
\sum_{j=1}^L\underline{\mathbf{H}}_{lk,j,i}
\underline{\mathbf{x}}_{j,i}
+
\underline{\mathbf{n}}_{lk,i},
\end{align}
where $\underline{\mathbf{n}}_{lk,i}$ is the real-decomposition of the effective additive zero-mean Gaussian noise at user $k$ in subband $i$ with covariance matrix $\underline{\mathbf{C}}_{lk,i}$, and $\underline{\mathbf{H}}_{lk,j,i}$ is the equivalent channel between BS $j$ and u$_{lk}$ at subband $i$, given by Lemma \ref{lem-iqi}. Note that each channel is a linear function of $\{\bm{\Phi}\}$, according to \eqref{ch-equ}. Hence, the effective channel is also a linear function of $\{\bm{\Phi}\}$. 
Moreover, note that all the signals $\underline{\mathbf{x}}_{lk,i}$s, are independent zero-mean and possibly improper Gaussian random vectors, where ${\bf P}_{lk,i}=\mathbb{E}\{\underline{\mathbf{x}}_{lk,i}\underline{\mathbf{x}}_{lk,i}^T\}$. 

The achievable rate of u$_{lk}$ is equal to 
\begin{equation}\label{rate-eq}
r_{lk}=\sum_{i=1}^{N_i}r_{lk,i},
\end{equation}
 where $r_{lk,i}$ is the rate of decoding $\mathbf{x}_{lk,i}$ at u$_{lk}$  treating intra-cell and inter-cell interference as noise 
\begin{subequations}\label{eq-28}
\begin{align}
r_{lk,i}
&=
\frac{1}{2}\log_2\left|
{\bf I}+
{\bf D}_{lk,i}^{-1}
 {\bf S}_{lk,i}
\right|
%r_{lk_1,i}\!-r_{lk_2,i},
\\ &
=
\underbrace{
\frac{1}{2}
\log_2
\left|{\bf D}_{lk,i}
+
{\bf S}_{lk,i}
\right|
}_{r_{lk,i_1}
}
-
\underbrace{
\frac{1}{2}\!\log_2\!\left|{\bf D}_{lk,i}
\right|}
_{r_{lk,i_2}
},
%\label{eq-29}
\end{align}
\end{subequations}
where ${\bf S}_{lk,i}=\underline{{\bf H}}_{lk,l,i}
{\bf P}_{lk,i}\underline{{\bf H}}_{lk,l,i}^T $ is the covariance matrix of the useful signal at u$_{lk}$ in subband $i$, and $\mathbf{D}_{lk,i}$ is the covariance matrix of noise plus interference  at u$_{lk}$ in subband $i$, given by
\begin{multline}
\mathbf{D}_{lk,i}
=
%\underbrace{
\sum_{n=1,n \neq l}^L
\underline{\mathbf{H}}_{lk,n,i}
\mathbf{P}_{n,i}\underline{\mathbf{H}}_{lk,n,i}^T
%}_{\text{Intercell interference}}%\!\!
\\
+
%\underbrace{
\sum_{m= 1,m\neq k}^{K}
\underline{\mathbf{H}}_{lk,l,i}
\mathbf{P}_{lm,i}
\underline{\mathbf{H}}_{lk,l,i}^T
%}_{\text{Intracell interference}}
+
%\underbrace{
\underline{\mathbf{C}}_{lk,i}.%}_{\text{Noise}},
\end{multline} 
%where $\underline{\mathbf{C}}_{i}$ is the noise covariance matrix at the receiver of u$_{lk}$ in subband $i$, given by Lemma \ref{lem-iqi}. 

The EE of u$_{lk}$ %a user is defined as the ratio between its rate and its total power consumption 
can be written as \cite{zappone2015energy}
\begin{equation}
e_{lk}=\frac{r_{lk}}{p_c+\eta\sum_{i}\text{Tr}\left(\mathbf{P}_{lk,i}\right)},
\end{equation} 
where %$\eta^{-1}$ is the power efficiency of each BS, and $p_c$ is the constant power consumption for transmitting data to a u$_{lk}$, given by
$\eta$ and $p_c$ are defined as in \cite[Eqs. (26)-(28)]{soleymani2022improper}.
Additionally, the global EE (GEE) can be obtained as %is defined as the total achievable rate (throughput) of the system, divided by the total power consumption, i.e., 
\cite{zappone2015energy}
\begin{equation}
GEE=\frac{\sum_{\forall l,k}r_{lk}}{LKp_c+\eta\sum_{l}\text{Tr}\left(\mathbf{P}_{l}\right)}, 
\end{equation}
where ${\bf P}_l=\sum_{k,i}{\bf P}_{lk,i}$.
% GEE is a metric for the performance of the whole network. Alternatively,  

\subsection{Problem statement}
%In this paper, w
We %consider 
aim at maximizing the SE and EE, which can be formulated %solving a general optimization problem 
as 
\begin{subequations}\label{ar-opt}
\begin{align}%\label{ar-opt}
 \underset{\{\mathbf{P}\},\{\bm{\Phi}\}\in\mathcal{T}
 }{\max}\,\,\,\,  & 
  f_0\!\left(\left\{\mathbf{P}\right\}\!,\!\{\bm{\Phi}\}\right)\\%!\!\!\! &
  \text{s.t.}  \,\,\, \,&  f_g\left(\left\{\mathbf{P}\right\}\!,\!\{\bm{\Phi}\}\right)\geq0,%\,\,\,\, 
&\forall g,
  \\
\label{25c} & \sum_{\forall k,i}\text{Tr}\left(\mathbf{P}_{lk,i}\right)\leq  P_l,&\forall l,
\\
\label{25d}&\mathbf{P}_{lk,i}\succcurlyeq\mathbf{0}, &\forall l,k,i
%\label{4-c}
% &&&
%r_{lk}\geq r^{th},\,\,\,\,\forall l,k,
 \end{align}
\end{subequations}
where $\left\{\mathbf{P}\right\}=\left\{\mathbf{P}_{lk,i}\right\}_{\forall l,k,i}$ and $\{\bm{\Phi}\}=\{\bm{\Phi}_m\}_{\forall m}$ are the optimization variables, and $P_l$ is the power budget of BS $l$. The functions $f_0$ and $f_i$ include EE and SE metrics, and thus, can be considered as a linear function of $r_{lk}$, and/or $e_{lk}$, and/or $GEE$. The optimization problems that can be cast as \eqref{ar-opt} have been discussed in \cite[Sec. II.D]{soleymani2022rate}. Such problems  include, but not limited to, the maximization of the minimum/sum rate, minimum EE and global EE. Note that these objective functions and/or optimization problems are among the most practical performance metrics in wireless communication systems.

\section{Optimization algorithms to solve \eqref{ar-opt}}\label{sec-iii}
To solve the non-convex problem \eqref{ar-opt}, we employ majorization minimization (MM) and alternating optimization (AO). To this end, we first fix the RIS coefficients to $\{\bm{\Phi}^{(t-1)}\}$, and solve \eqref{ar-opt} to obtain $\left\{\mathbf{P}^{(t)}\right\}$. We then fix the transmit covariance matrices to $\left\{\mathbf{P}^{(t)}\right\}$ and update $\{\bm{\Phi}\}$ by solving \eqref{ar-opt}. 
In the following, we describe our proposed algorithm to update $\left\{\mathbf{P}\right\}$ and  $\bm{\Phi}$ in separate subsections.  

\subsection{Updating transmit covariance matrices} 
When RIS components are fixed to $\bm{\Phi}^{(t-1)}$, \eqref{ar-opt} is  
\begin{subequations}\label{ar-opt-p}
\begin{align}%\label{ar-opt-p}
 \underset{\{\mathbf{P}\}%\in\mathcal{P}
 }{\max}\,\,\,  & 
  f_0\left(\left\{\mathbf{P}\right\}\!,\!\{\bm{\Phi}^{(t-1)}\}\right)\\
  \text{s.t.}   \,\,\,&  f_g\left(\left\{\mathbf{P}\right\}\!,\!\{\bm{\Phi}^{(t-1)}\}\right)\geq0,\,\forall g,
  \\ & \eqref{25c},\eqref{25d},
 \end{align}
\end{subequations}
which is non-convex since $r_{lk}\left(\left\{\mathbf{P}\right\}\!,\!\bm{\Phi}^{(t-1)}\right)$ for all $l,k$ is not concave in $\left\{\mathbf{P}\right\}$. 
%We employ an MM-based scheme t
To provide a suboptimal solution for \eqref{ar-opt-p}, we obtain suitable concave lower bounds for $r_{lk}\left(\left\{\mathbf{P}\right\}\!,\!\{\bm{\Phi}^{(t-1)}\}\right)$. %, which results in a convex surrogate optimization problem. 
To this end, we employ the lower bound in \cite[Lemma 3]{soleymani2022rate}, which results in the concave lower bounds for the rates that are provided in Lemma \ref{coro-1}. 
\begin{lemma}%{corollary}
 \label{coro-1}
A concave lower bound for $r_{lk}$ is $\bar{r}_{lk}=\sum_{i}\bar{r}_{lk,i}$, where $\tilde{r}_{lk,i}$ is 
\begin{multline*}%{equation}
%\label{l-r-lk-p}
 r_{lk,i}\geq \bar{r}_{lk,i}=
r_{lk,i_1}\left(\{\mathbf{P}\}\right) 
-r_{lk,i_2}^{(t-1)}
\\
-\!\!\!
\sum_{j=1,\neq k}^{K}\!\!\!\!
\mathfrak{R}\!\left\{\!
\text{\em Tr}\!\left(\!
\frac{
%\underline
{\mathbf{H}}_{lk,l,i}^H
(\mathbf{D}_{lk,i}^{(t-1)})^{-1}
{\mathbf{H}}_{lk,l,i}
}
{\ln 2}
\left(\mathbf{P}_{lj,i}-\mathbf{P}_{lj,i}^{(t-1)}\right)\!\!
\right)\!\!
\right\}
\\
-\!\!
\sum_{n=1,\neq l}^{L}\!\!\!
\mathfrak{R}\!\left\{\!
\text{\em Tr}\!\left(\!
\frac{
%\underline
{\mathbf{H}}_{lk,n,i}^H
(\mathbf{D}_{lk,i}^{(t-1)})^{-1}
{\mathbf{H}}_{lk,n,i}
}
{\ln 2}\!
\left(\mathbf{P}_{n,i}-\mathbf{P}_{n,i}^{(t-1)}\right)\!\!
\right)\!\!
\right\}\!
,
\end{multline*}
where $r_{lk,i_2}^{(t-1)}=r_{lk,i_2}\left(\{\mathbf{P}^{(t-1)}\}\right)$, and 
$\mathbf{D}_{lk,i}^{(t-1)}=\mathbf{D}_{lk,i}\left(\{\mathbf{P}^{(t-1)}\}\right)$.
\end{lemma}%{corollary}
Substituting ${r}_{lk}$ with $\bar{r}_{lk}$, 
we have
\begin{subequations}\label{ar-opt-p-sur}
\begin{align}%\label{ar-opt-p}
 \underset{\{\mathbf{P}\}%\in\mathcal{P}
 }{\max}\,\,\,  & 
  \bar{f}_0\left(\left\{\mathbf{P}\right\}\!,\!\{\bm{\Phi}^{(t-1)}\}\right)\\
  \text{s.t.}   \,\,\,&  \bar{f}_g\left(\left\{\mathbf{P}\right\}\!,\!\{\bm{\Phi}^{(t-1)}\}\right)\geq0,\,\forall g,
\\ & \eqref{25c},\eqref{25d},
 \end{align}
\end{subequations}
which is convex in $\{{\bf P}\}$ when SE metrics such as minimum and/or sum rate are considered. For EE metrics such as GEE and/or minimum weighted EE, we can employ Dinkelbach-based algorithms to obtain a global optimal solution of \eqref{ar-opt-p-sur}. Note that the solution of \eqref{ar-opt-p-sur} forms the new set of the transmit covariance matrices, i.e., $\left\{\mathbf{P}^{(t)}\right\}$, which is utilized in the next step.

\subsection{Updating RIS elements}
In this subsection, we first propose optimization algorithms to update $\{\bm{\Phi}\}$ for the ES scheme since it is more general than the MS scheme. Indeed,  the MS scheme can be considered as a special case of the ES scheme as discussed in Section \ref{seciia}, which means that the solutions for the ES scheme can be applied to the MS scheme as well. At the end of this subsection, we explain how the ES solutions can be modified to include the MS scheme.
 
When transmit covariance matrices are fixed to $\left\{\mathbf{P}^{(t)}\right\}$, \eqref{ar-opt} is equivalent to 
\begin{subequations}\label{ar-opt-t}
\begin{align}%\label{ar-opt}
 \underset{\{\bm{\Phi}\}\in\mathcal{T}
 }{\max}\,\,\,\,  & 
  f_0\!\left(\left\{\mathbf{P}^{(t)}\right\}\!,\!\{\bm{\Phi}\}\right)\\%!\!\!\! &
  \text{s.t.}  \,\,\, \,&  f_g\left(\left\{\mathbf{P}^{(t)}\right\}\!,\!\{\bm{\Phi}\}\right)\geq0,\,\,\,\,\forall g,
   \end{align}
\end{subequations}
which is non-convex since $r_{lk}\left(\left\{\mathbf{P}^{(t)}\right\}\!,\!\{\bm{\Phi}\}\right)$ is not concave in $\{\bm{\Phi}\}$, and moreover,  $\mathcal{T}$ is not a convex set for $\mathcal{T}_I$ and $\mathcal{T}_{SN}$. 
To provide a suboptimal solution for \eqref{ar-opt-t}, we first obtain a concave lower bound for  $r_{lk}\left(\left\{\mathbf{P}^{(t)}\right\}\!,\!\{\bm{\Phi}\}\right)$, and then convexify $\mathcal{T}_I$, and/or $\mathcal{T}_{SN}$. %if it is not a convex set.
 To this end, we employ the bound in \cite[Lemma 2]{soleymani2022improper}, which gives the surrogate functions for the rates as in the lemma below. 
\begin{lemma}%{corollary} 
\label{theo1}
A concave lower-bound for the rate of users is $\hat{r}_{lk}=\sum_i\hat{r}_{lk,i}$, where $\hat{r}_{lk,i}$ is %a concave lower bound for ${r}_{lk,i}$ as
%\begin{figure*}[t]
\begin{multline} 
%\nonumber
r_{lk,i}\geq
 \hat{r}_{lk,i}=r_{lk,i}^{(t-1)}-
\frac{1}{\ln 2}\left(\text{{\em Tr}}\left(
\bar{\mathbf{S}}_{lk,i}\bar{\mathbf{D}}_{lk,i}^{-1}
\right)\right.
\\
-
\text{{\em Tr}}\left(
(\bar{\mathbf{D}}^{-1}_{lk,i}\!-\!(\bar{\mathbf{S}}_{lk,i} + \bar{\mathbf{D}}_{lk,i})^{-1})^H (\mathbf{S}_{lk,i}+\mathbf{D}_{lk,i})
\right)
\\
\left.
+
2\mathfrak{R}\left\{
\text{{\em Tr}}\left(
\bar{\mathbf{V}}_{lk,i}^H\bar{\mathbf{D}}_{lk,i}^{-1}\mathbf{V}_{lk,i}
\right)\right\}
\right),
%\label{lower-bound-th1}
\end{multline}
%\hrulefill %\vspace{-.4cm}
%\end{figure*}
where $r_{lk,i}^{(t-1)}=r_{lk,i}\left(%\{\mathbf{P}^{(t)}\},
\{\bm{\Phi}^{(t-1)}\}\right)$, %and 
$%\begin{align*}
\mathbf{V}_{lk,i}%&
=%\underline
{\mathbf{H}}_{lk,l,i}\left(\bm{\Phi}\right)\mathbf{P}_{lk,i}^{(t)^{1/2}}$,
%& \hspace{1cm}
%\\
%, 
$\bar{\mathbf{D}}_{lk,i}=\mathbf{D}_{lk,i}\left(\{\bm{\Phi}^{(t-1)}\}\right)$, 
$\bar{\mathbf{S}}_{lk,i}=\mathbf{S}_{lk,i}\left(\{\bm{\Phi}^{(t-1)}\}\right)$, and 
$\bar{\mathbf{V}}_{lk,i}=%\underline
{\mathbf{H}}_{lk,l,i}\left(\{\bm{\Phi}^{(t-1)}\}\right)\mathbf{P}_{lk,i}^{(t)^{1/2}}$.
\end{lemma}%{corollary}
Substituting ${r}_{lk}$ with $\hat{r}_{lk}$, \eqref{ar-opt-t} becomes %the following surrogate problem
\begin{subequations}\label{ar-opt-t-sur}
\begin{align}%\label{ar-opt}
 \underset{\{\bm{\Phi}\}\in\mathcal{T}
 }{\max}\,\,\,\,  & 
  \hat{f}_0\!\left(\left\{\mathbf{P}^{(t)}\right\}\!,\!\{\bm{\Phi}\}\right)\\%!\!\!\! &
  \text{s.t.}  \,\,\, \,&  \hat{f}_g\left(\left\{\mathbf{P}^{(t)}\right\}\!,\!\{\bm{\Phi}\}\right)\geq0,\,\,\,\,\forall g,
 \end{align}
\end{subequations}
which is convex  only for $\mathcal{T}_U$ in which the convergence to a stationary point of \eqref{ar-opt} is ensured. However, \eqref{ar-opt-t-sur} is non-convex for $\mathcal{T}_I$ (or $\mathcal{T}_{SN}$) because of the constraint in \eqref{sr-2} (or \eqref{st-eq-2}). 
%To find a suboptimal solution for   
We can rewrite $\sum_{n_s=1}^{N_s}|\phi_{mn}^{n_s}|^2=1$ as the two constraints %\eqref{sr-1} and %the two following constraints
\begin{align}\label{14}
\sum_{n_s=1}^{N_s}|\phi_{mn}^{n_s}|^2&\leq 1 \hspace{1cm}\forall m,n,
\\
\sum_{n_s=1}^{N_s}|\phi_{mn}^{n_s}|^2&\geq 1 \hspace{1cm}\forall m,n,
\label{15}
\end{align}
The constraint \eqref{14} is convex. However, \eqref{15} is a non-convex constraint, which makes \eqref{ar-opt-t-sur} non-convex for $\mathcal{T}_I$ (or $\mathcal{T}_{SN}$). To approximate \eqref{15} with a convex constraint, we can employ the convex-concave procedure (CCP) and relax the constraint for a  faster convergence as \cite{soleymani2022improper}
\begin{equation}\label{17}
\sum_{n_s=1}^{N_s}\!\!\left(\!\!2\mathfrak{R}\left\{\!
\phi_{mn}^{n_s^{(t-1)^*}}\phi_{mn}^{n_s}\!\right\}-
|\phi_{mn}^{n_s^{(t-1)}}|^2\!
\right)\!\!
\geq \!1-\epsilon, \hspace{.15cm} \forall m,n,
\end{equation}
where $\epsilon>0$. Substituting the constraints \eqref{14} and \eqref{17} in \eqref{ar-opt-t-sur} yields %the following convex problem
\begin{subequations}\label{ar-opt-18}
\begin{align}%\label{ar-opt-t-sur}
 \underset{\{\bm{\Phi}\}
 }{\max}\,\,\,\,\,  & 
  \hat{f}_0\!\left(\left\{\mathbf{P}^{(t)}\right\}\!,\!\{\bm{\Phi}\}\right)\\%!\!\!\! &
  \text{s.t.}  \,\,\,\,\, \,&  \hat{f}_g\left(\left\{\mathbf{P}^{(t)}\right\}\!,\!\{\bm{\Phi}\}\right)\geq0,\,\,\,\,\forall g\\
%&&
&\eqref{14},\eqref{17},
 \end{align}
\end{subequations}
which can be efficiently solved since \eqref{ar-opt-18} is convex. 
We denote the solution of \eqref{ar-opt-18} as $\{\bm{\Phi}^{(\star)}\}$. 
It might happen that $\{\bm{\Phi}^{(\star)}\}$ does not meet the  constraint in \eqref{sr-2} due to the relaxation in \eqref{17}. 
To generate a feasible solution, we normalize $\{\bm{\Phi}^{(\star)}\}$ as 
\begin{equation}
\hat{\phi}_{mn}^{n_s}=\frac{\phi^{n_s^{(\star)}}_{mn}}{\sqrt{\sum_{n_s=1}^{N_s}\left|\phi^{n_s^{(\star)}}_{mn}\right|^2}},\hspace{1cm}\forall m,n.
\end{equation}
Finally, we update $\{\bm{\Phi}\}$ as
\begin{equation}\label{eq-19}
\{\bm{\Phi}^{(t)}\}\!\!=\!\!
\left\{\!\!\!\!
\begin{array}{lcl}
\{\hat{\bm{\Phi}}\}\!\!&\!\!\!\!\!\!\text{if}\!\!\!\!\!&
\hat{f}_0\!\left(\left\{\mathbf{P}^{(t)}\right\}\!,\!\{\hat{\bm{\Phi}}\}\right)
\geq
\\
&&
\hat{f}_0\!\left(\left\{\mathbf{P}^{(t)}\right\}\!,\!\{\bm{\Phi}^{(t-1)}\}\right)
\\
\{\bm{\Phi}^{(t-1)}\}&&\text{otherwise},
\end{array}
\right.
\end{equation}
where $\{\hat{\bm{\Phi}}\}=\{\hat{\bm{\Phi}}_m^{n_s}\}_{\forall n_s,m}$, where 
\begin{equation}\hat{\bm{\Phi}}_m^{n_s}=\text{diag}(\hat{\phi}_{m1}^{n_s},\hat{\phi}_{m2}^{n_s},\cdots,\hat{\phi}_{m_{N_\mathsf{RIS}}}^{n_s}).
\end{equation}
The updating policy in \eqref{eq-19} ensures convergence because of generating a non-decreasing sequence of $\hat{f}_0(\cdot)$.

Note that for the set $\mathcal{T}_{SN}$, we can employ a similar approach since the constraints \eqref{st-eq-1} and \eqref{st-eq-} are convex, and we can handle \eqref{st-eq-2} similar to $\sum_{n_s=1}^{N_s}|\phi_{mn}^{n_s}|^2=1$ for $N_s=2$. Hence, to update $\{\bm{\phi}\}$, we have to solve the convex problem
\begin{subequations}\label{ar-opt-18+}
\begin{align}%\label{ar-opt-t-sur}
 \underset{\{\bm{\phi}\}
 }{\max}\,\,\,\,\,  & 
  \hat{f}_0\!\left(\left\{\mathbf{P}^{(t)}\right\}\!,\!\{\bm{\Phi}\}\right)\\%!\!\!\! &
  \text{s.t.}  \,\,\,\,\, \,&  \hat{f}_g\left(\left\{\mathbf{P}^{(t)}\right\}\!,\!\{\bm{\Phi}\}\right)\geq0,\,\,\,\,\forall g\\
%&&
&\eqref{st-eq-1},\eqref{st-eq-},\eqref{14},\eqref{17},
 \end{align}
\end{subequations}
%which is a convex optimization problem. 
Then we update $\{\bm{\Phi}\}$ according to the rule in \eqref{eq-19}.

Now we present our solution for the MS scheme. That is,
we randomly divide the multi-sector BD-RIS elements into $N_s$ groups such that there are at least $\lfloor\frac{N_{RIS}}{N_s}\rfloor$ elements per group. 
Then the RIS elements in the $n_s$-th group, indicated by $\mathcal{G}_{mn_s}$, operate only in the sector $n_s$, and thus, $\phi_{mn}^{n_s}=0$ if $n\notin\mathcal{G}_{mn_s}$. 
Inserting this rule in \eqref{ar-opt-t-sur} and \eqref{ar-opt-18} gives the MS solution for $\mathcal{T}_U$ and $\mathcal{T}_I$, respectively. For $\mathcal{T}_I$, we have to update $\{\bm{\Phi}\}$ based on the rule in \eqref{eq-19}. 
Note that the sets $\mathcal{T}_I$ and $\mathcal{T}_{SN}$ are equivalent for MS schemes since the phase dependency constraints are automatically satisfied when each STAR-RIS element is active in only one sector.

\section{Numerical results}\label{sec-nr}
In this section, we employ Monte Carlo simulations to provide numerical results. We assume that the small-scale fading for the channels ${\bf G}_i$ and ${\bf G}_{ki}$ for all $i,k$ is Rician  similar to \cite{pan2020multicell, soleymani2022rate} since there is a line of sight (LoS) link between the BS and RIS as well as between the RIS and the users. However, as  the links between the BS and the users are assumed to be non-LoS (NLoS), the small-scale fading for ${\bf F}_{ki}$ for all $k,i$ is assumed to be Rayleigh distributed. 
Note that it is also assumed that there is no correlation between the channels at different subbands, which is an extreme case, and it can be expected that the performance of RIS is improved with correlated channels. 
The simulation scenario is based on \cite[Fig. 2]{soleymani2022rate}, unless explicitly mentioned otherwise. For more descriptions on the simulation parameters and setup, we refer the reader to \cite{soleymani2022rate, soleymani2022noma}.

\subsection{Maximization of the minimum rate }
Here, we investigate the effectiveness of the various RIS technologies from a SE point of view by considering the maximization of the minimum rate. To this end, we consider the impact of different system parameters, including the power budget at the BSs, the number of subbands, the number of RIS elements, and IQI parameters. %on the performance of RIS.

\subsubsection{Impact of power budget}
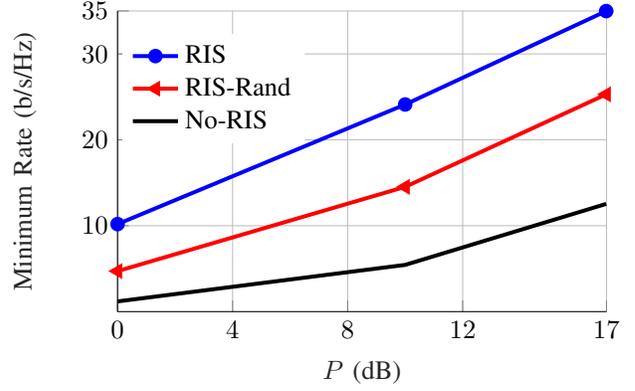
\begin{figure}[t!]
       \input{r}
    \caption{Average minimum rate versus $P$ for the case without RIS (No-RIS), randomly configured RIS (RIS-Rand), and optimized regular RIS (RIS) with $N_{\mathsf{B}}=N_{\mathsf{U}}=2$, $K=3$, $N_{\mathsf{R}}=100$,   $L=2$,  $N=2$, and $N_i=16$.}
	\label{Fig-rr1} 
\end{figure}
Fig. \ref{Fig-rr1} shows the average minimum rate versus $P$ for $N_{\mathsf{B}}=N_{\mathsf{U}}=2$, $K=3$, $N_{\mathsf{R}}=100$,  $L=2$,  $N=2$, and $N_i=16$. In this figure, we can observe that the regular RIS can substantially increase the minimum rate even when the RIS elements are not optimized (RIS-Rand).
Additionally, there is almost a constant gap between the proposed scheme for RIS-aided systems and the scheme with random RIS coefficients (RIS-Rand). Indeed, even though the RIS elements cannot be independently optimized at each subband, we can get a significant gain by optimizing RIS coefficients, which shows the effectiveness of RIS in multi-user MIMO OFDM systems.

\subsubsection{Impact of the number of subbands}
\begin{figure}[t!]
    \centering
    \begin{subfigure}[t]{0.24\textwidth}
        \centering
       \includegraphics[width=\textwidth]{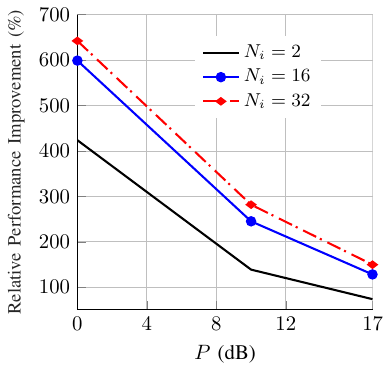}%{con2}
        \caption{RIS compared to No-RIS.}
    \end{subfigure}%
    ~
    \begin{subfigure}[t]{0.24\textwidth}
        \centering
           \includegraphics[width=\textwidth]{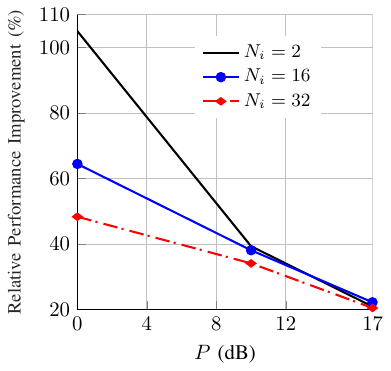}%{con3}
        \caption{RIS compared to RIS-Rand.}
    \end{subfigure}
    \caption{Improvements by RIS versus $P$ for $N_{\mathsf{B}}=2$, $N_{\mathsf{U}}=2$, $K=3$,  $L=1$,  $N=1$, and different $N_i$.}
	\label{Fig-rr2} 
\end{figure}
Fig. \ref{Fig-rr2} shows the average performance improvements versus $P$ for $N_{\mathsf{B}}=N_{\mathsf{U}}=2$,  $K=3$, $N_{\mathsf{R}}=100$,  $L=1$,  $N=1$, different $N_i$ and $a_{t,i}=a_{r,i}=1$ for all $i$. The relative performance curves in Fig. \ref{Fig-rr2}a and Fig. \ref{Fig-rr2}b are obtained by comparing the average minimum rate achieved by our scheme for RIS-aided OFDM systems with the average minimum rate of OFDM systems without RIS and with the average minimum rate of OFDM systems with random RIS coefficients, respectively. According to Fig. \ref{Fig-rr2}, RIS can provide a huge gain. However, the benefits of optimizing RIS components highly decrease with $N_i$ since RIS elements cannot be independently optimized at each subband, and as the number of subbands for a fixed $N_{\mathsf{R}}$ increases, the effectiveness of optimizing $\bm{\Phi}$ decreases.  
Interestingly, the benefits of optimizing RIS elements are still significant even when there are slightly higher than $1$ RIS elements per user per subband ($N_{\mathsf{R}}/(KN_i)\simeq 1.04$ when $N_i$=32).  

Furthermore, the gains of optimizing RIS elements are much higher in low SNR regimes. 
Since we also consider a power/covariance matrix optimization, it may happen that the signals for a user are transmitted over a few subbands when the BS power budget is low. Thus, the effective number of the utilized subbands is lower than $N_i$, especially at low SNR regimes, which enhances the benefits of optimizing RIS elements.  Additionally, as shown in our previous studies \cite{soleymani2022improper,soleymani2022rate}, the benefits of employing RIS are higher in low SNR regimes, which may enhance the gain of a proper optimization of RIS elements.

\subsubsection{Impact of $N_{\mathsf{R}}$}
\begin{figure}[t!]
    \centering
    \begin{subfigure}[t]{0.24\textwidth}
        \centering
       \includegraphics[width=\textwidth]{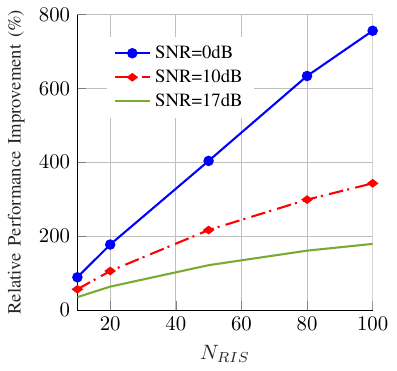}
        \caption{RIS compared to No-RIS.}
    \end{subfigure}%
    %\\
    \begin{subfigure}[t]{0.24\textwidth}
        \centering
           \includegraphics[width=\textwidth]{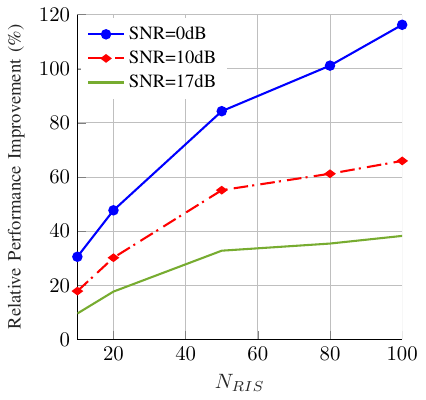}
        \caption{RIS compared to RIS-Rand.}
    \end{subfigure}
    \caption{Improvements by RIS versus $P$ for $N_{\mathsf{B}}=2$, $N_{\mathsf{U}}=2$, $K=3$,  $L=2$,  $N=2$, and $N_i=16$.}
	\label{Fig-rr3} 
\end{figure}
Fig. \ref{Fig-rr3} shows the average performance improvements versus $N_{\mathsf{R}}$ for $N_{\mathsf{B}}=N_{\mathsf{U}}=1$, $K=3$, $L=1$, $N_i=16$, $N=1$, and different SNRs. In this figure, the benefits of employing RIS almost linearly increases with $N_{\mathsf{R}}$, especially at low SNR. Additionally, we can observe that the slope of the RIS benefits is higher in lower SNR, which is in line with the results in the Figs. \ref{Fig-rr1} and \ref{Fig-rr2}.
 Moreover,  optimizing  RIS elements can yield substantial gains even when $N_{\mathsf{R}}=10$, which is lower than the number of sub-bands (in this case $N_{\mathsf{R}}/(KN_i)\simeq 0.21$).

\subsubsection{Impact of IQI}
\begin{figure}[t!]
       \includegraphics[width=.45\textwidth]{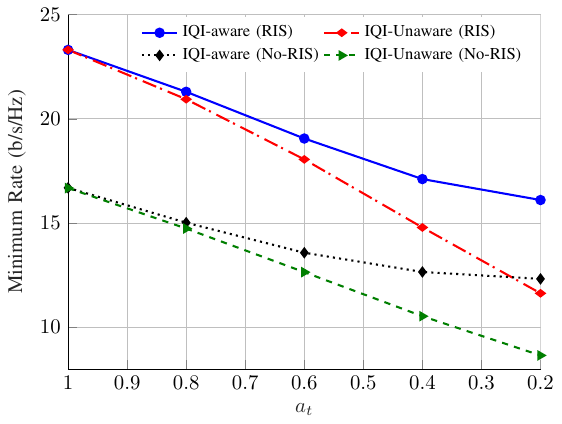}
    \caption{Impact of IQI on the average minimum rate of users for $N_{\mathsf{B}}=2$, $N_{\mathsf{U}}=2$, $K=2$, $N_{\mathsf{R}}=100$,   $L=2$,  $N=2$, and $N_i=16$.}
	\label{Fig-rr4} 
\end{figure}
Fig. \ref{Fig-rr4} shows the average minimum rate of users versus $a_t$ for $N_{\mathsf{B}}=2,$ $N_{\mathsf{U}}=2$, $K=2$, $N_{\mathsf{R}}=100$,   $L=2$,  $N=2$, and $N_i=16$. In this figure, we assume that both transceivers suffer from IQI with the same IQI parameters at all subbands, which means that $a_{r,i}=a_{t,i}=a_t$ for all $i$.  This figure shows that the average minimum rate decreases with IQI level at transceivers even if it is considered in the system designed and compensated by IGS. 
Furthermore, we can observe that IQI can significantly decrease the average minimum rate 
especially when we do not take it into account in optimizing the parameters.
Indeed, IQI may even vanish the benefits of RIS in systems with highly imbalanced devices if the employed schemes are not robust against IQI. 
Additionally, we observe that IQI has a similar impact on the systems with and without RIS; however, the benefits of IQI-aware schemes are a bit higher in RIS-aided systems. 

\subsubsection{Comparison of different technologies for RIS}
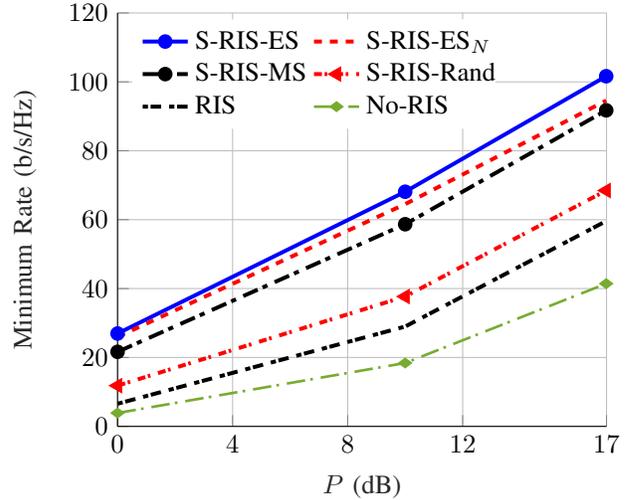
\begin{figure}[t!]
       \input{r-star}
    \caption{Average minimum rate versus $P$ for the case without RIS (No-RIS), randomly configured STAR-RIS (S-RIS-Rand), optimized STAR-RIS (S-RIS) with different operational mode and feasibility sets, and optimized regular RIS (RIS) for $N_{\mathsf{B}}=N_{\mathsf{U}}=2$, $K=2$, $N_{\mathsf{R}}=100$,   $L=1$,  $N=1$, and $N_i=16$.}
	\label{Fig-rr5} 
\end{figure}
Here, we assume $L=1$ and $N=1$, which means that we consider only one RIS (either multi-sector BD-RIS, STAR or regular). To show the impact of an omin-directional coverage, we consider a scenario in which the users are located such that the regular RIS can cover only  half of the users, while the STAR-RIS and/or multi-sector BD-RIS  can cover all the users.

In Fig. \ref{Fig-rr5}, we show the average minimum rate versus $P$  for $N_{\mathsf{B}}=N_{\mathsf{U}}=2$, $K=2$, $N_{\mathsf{R}}=100$,   $L=1$,  $N=1$, and $N_i=16$. 
This figure compares the performance of a regular RIS and a STAR-RIS by considering different feasibility sets as well as different operational modes for the STAR-RIS. 
In this example, RIS (either regular or STAR) can highly increase the average minimum rate. Additionally, a STAR-RIS with even random coefficients can outperform the proposed scheme for the regular RIS.  Moreover,  optimizing the STAR-RIS coefficients can significantly increase the average minimum rate.
Furthermore, the ES scheme outperforms the MS scheme; however, considering the higher computational and implementation complexities of ES schemes, the MS scheme performance is comparable with the ES scheme, especially for  $\mathcal{T}_{SN}$ in which the phases of transmit and reflection coefficients cannot be independently optimized.

\begin{figure}[t!]
    \centering
    \begin{subfigure}[t]{0.24\textwidth}
        \centering
           \includegraphics[width=\textwidth]{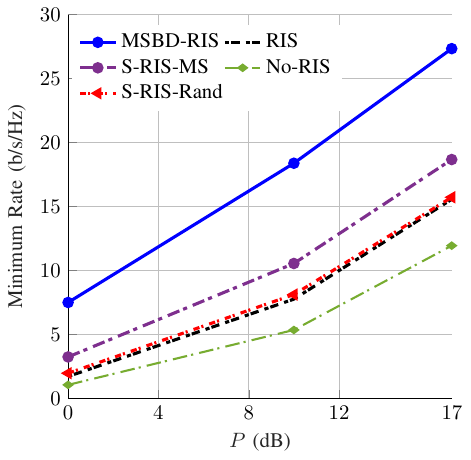}
        \caption{$N_{\mathsf{B}}=N_{\mathsf{U}}=1$.}
    \end{subfigure}
\begin{subfigure}[t]{0.24\textwidth}
        \centering
       \includegraphics[width=\textwidth]{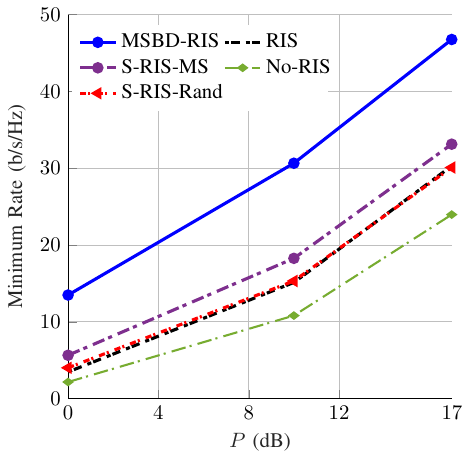}
        \caption{$N_{\mathsf{B}}=N_{\mathsf{U}}=2$.}
    \end{subfigure}%
    \caption{The average minimum rate versus $P$ for the case without RIS (No-RIS), randomly configured STAR-RIS (S-RIS-Rand), optimized STAR-RIS with mode switching scheme (S-RIS-MS), optimized regular RIS (RIS), and optimized multi-sector BD-RIS with mode switching scheme (MSBD-RIS) for $K=4$, $N_{\mathsf{R}}=32$,   $L=1$,  $N=1$, and $N_i=16$.}
	\label{Fig-rr6} 
\end{figure}
Fig. \ref{Fig-rr6} compares the performance of various RIS technologies from an average minimum rate point of view for $K=4$, $N_{\mathsf{R}}=32$,   $L=1$,  $N=1$, $N_i=16$, and different $N_{\mathsf{B}}$, $N_{\mathsf{U}}$. To this end, we consider a four-sector BD-RIS ($N_s=4$), a STAR-RIS and a regular RIS. We assume that there is one user in each sector of the multi-sector BD-RIS. Moreover, to make the comparison fair, we assume that the total number of the RIS elements is the the same and is equal to $32$ in these different technologies. 
It means that there are only $N_{\mathsf{R}}/N_s$ elements per each sector of the multi-sector BD-RIS, which is equal to $8$ in this example. 
Note that we assume that STAR-RIS has $N_{\mathsf{R}}$ (32 in this example) elements since STAR-RIS can employ a different technology comparing to multi-sector BD-RIS. 
Furthermore, we assume that the antenna gain of each multi-sector BD-RIS  is computed based on the idealized model in \cite[Eq. (17)]{li2023beyond}. Here, we consider only the MS scheme since it has lower computational and implementation complexities while providing a competitive performance, as shown in Fig. \ref{Fig-rr5}. 

In Fig. \ref{Fig-rr6}, all the considered RIS technologies can substantially increase the average minimum rate with even a very low $N_{\mathsf{R}}$. In this example, the number of RIS elements per user per subband is $0.5$, which is even lower than 1. We can also observe that STAR-RIS can highly outperform a regular RIS. Additionally,  the multi-sector BD-RIS can provide a significant gain even though the number of RIS elements per sector is much lower than the number of STAR-RIS elements. 

\begin{figure}[t!]
       \includegraphics[width=.4\textwidth]{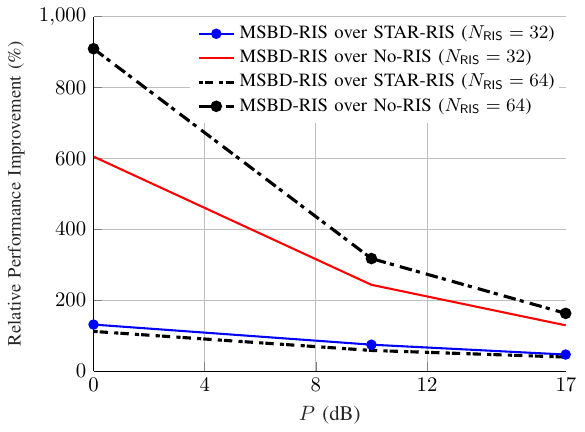}
    \caption{The average improvements by multi-sector BD-RIS versus $P$ for  $N_{\mathsf{B}}=N_{\mathsf{U}}=1$, $K=4$,    $L=1$,  $N=1$, $N_i=16$, and $N_{\mathsf{R}}$.}
	\label{Fig-rr7} 
\end{figure}
Fig. \ref{Fig-rr7} shows the average improvements by multi-sector BD-RIS over STAR-RIS and/or No-RIS versus $P$ for  $N_{\mathsf{B}}=1$, $N_{\mathsf{U}}=1$, $K=4$,    $L=1$,  $N=1$, $N_i=16$, and $N_{\mathsf{R}}$. 
As can be observed, multi-sector BD-RIS can substantially improve the SE, especially in low SNR regimes. Additionally, the benefits of employing RIS increase with $N_{\mathsf{R}}$.

\subsection{Maximization of the sum rate}
\begin{figure}[t!]
    \centering
    \begin{subfigure}[t]{0.24\textwidth}
        \centering
           \includegraphics[width=\textwidth]{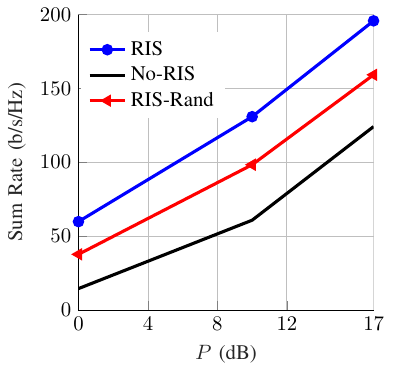}
        \caption{Average sum rate.}
    \end{subfigure}
\begin{subfigure}[t]{0.24\textwidth}
        \centering
        \includegraphics[width=\textwidth]{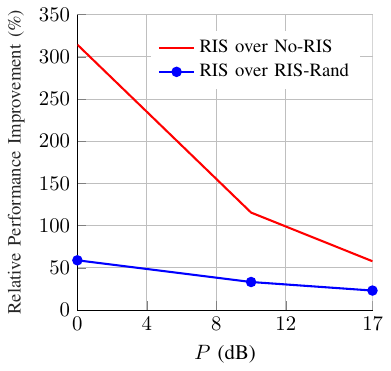}
        \caption{Relative performance improvement.}
    \end{subfigure}%
    \caption{The average sum rate versus $P$ for $N_{\mathsf{B}}=2$, $N_{\mathsf{U}}=2$, $K=3$, $L=2$,  $N=2$, $N_i=16$, and $N_{\mathsf{R}}=50$.}
	\label{Fig-sr} 
\end{figure}
Fig. \ref{Fig-sr} shows the average sum rate versus $P$ for $N_{\mathsf{B}}=N_{\mathsf{U}}=2$, $K=3$, $L=2$,  $N=2$, $N_i=16$, and $N_{\mathsf{R}}=50$.
In this example, RIS can substantially increase the average sum rate. 
Moreover, the benefits of RIS decrease with the power budget, which supports the results in Fig. \ref{Fig-rr1}. Indeed, in low SNR regimes, it is likely that the BS transmits over a few number of subbands, which can increase the benefits of optimizing RIS elements since the RIS coefficients cannot be independently optimized at each subband.
Additionally, as shown in, e.g., \cite{soleymani2022improper}, RIS can provide higher gains in low SNR regimes in single-carrier systems since the received power of cell-edge users is very low in systems without RIS when the BS transmission power is low. Hence, we can  expect that this result hold in multi-carrier case, which highly increases the benefits of RIS in low SNR regimes.  

\subsection{Maximization of the minimum EE}
\begin{figure}[t!]
    \centering
    \begin{subfigure}[t]{0.24\textwidth}
        \centering
           \includegraphics[width=\textwidth]{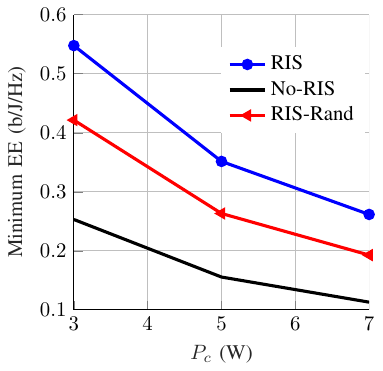}
        \caption{Average minimum EE.}
    \end{subfigure}
\begin{subfigure}[t]{0.24\textwidth}
        \centering
       \includegraphics[width=\textwidth]{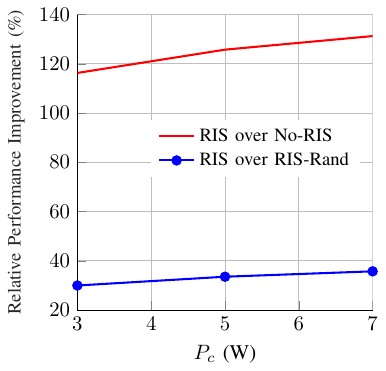}
        \caption{Relative performance improvement.}
    \end{subfigure}%
    \caption{The average minimum EE versus $P_c$ for $N_{\mathsf{B}}=N_{\mathsf{U}}=2$, $K=2$, $N_{\mathsf{R}}=50$, $L=2$,  $N=2$, and $N_i=16$.}
	\label{Fig-ee} 
\end{figure}
Fig. \ref{Fig-ee} shows the average minimum EE versus $P_c$ for $N_{\mathsf{B}}=N_\mathsf{U}=2$, $K=2$, $N_{\mathsf{R}}=50$, $L=2$,  $N=2$, and $N_i=16$. In this example, RIS can significantly improve the average minimum EE of OFDM systems with different $P_c$. However, the EE benefits of RIS are lower than its SE benefits. Additionally,  the RIS benefits  slightly increase with $P_c$. The reason is that, when $P_c$ asymptotically increases, the solutions of the maximization of the minimum rate and the maximization of the minimum EE are identical since the EE maximization for a large $P_c$ is equivalent to the rate maximization. Hence, as $P_c$ increases, the EE benefits of RIS become closer to the SE benefits, which makes the EE benefits of RIS an increasing function of $P_c$.  

\subsection{Maximization of the global EE}
\begin{figure}[t!]
\includegraphics[width=.4\textwidth]{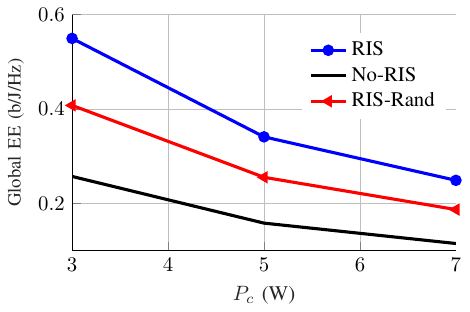}
    \caption{The average global EE versus $P_c$ for $N_{\mathsf{B}}=2$, $N_\mathsf{U}=2$, $K=2$, $N_{\mathsf{R}}=40$, $L=2$,  $N=2$, and $N_i=16$.}
	\label{Fig-gee} 
\end{figure}
Fig. \ref{Fig-gee} shows the average global EE versus $P_c$ for $N_{\mathsf{B}}=2$, $N_\mathsf{U}=2$, $K=2$, $N_{\mathsf{R}}=40$, $L=2$,  $N=2$, and $N_i=16$.
In this figure, RIS can significantly enhance the global EE of the OFDM system for different $P_c$. Moreover, optimizing RIS elements provides more that $32\%$ improvements in the particular example with a relatively low number of RIS elements per user per subband ($N_{\mathsf{R}}/(KN_i)\simeq 1.25$).

\begin{figure}[t!]
\includegraphics[width=.4\textwidth]{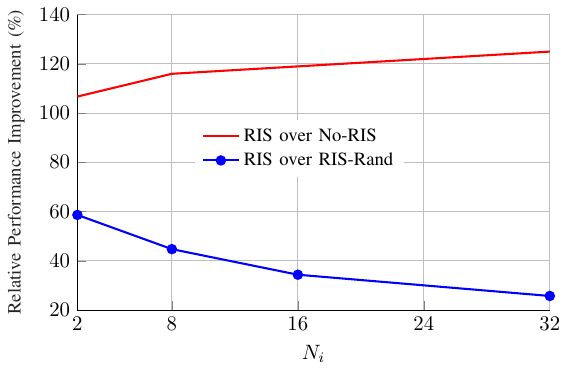}
    \caption{The average benefits of RIS from a global EE point of view versus $N_i$ for $N_{\mathsf{B}}=2$, $N_\mathsf{U}=2$, $K=2$, $N_{\mathsf{R}}=40$, $L=2$,  $N=2$, and $P_c=5$ W.}
	\label{Fig-gee2} 
\end{figure}
Fig. \ref{Fig-gee2} shows the average benefits of RIS from a global EE point of view versus $N_i$ for $N_{\mathsf{B}}=2$, $N_\mathsf{U}=2$, $K=2$, $N_{\mathsf{R}}=40$, $L=2$,  $N=2$, and $P_c=5$ W.
As can be observed, %similar to the results in Fig. \ref{Fig-rr3}, 
the benefits of employing RIS (compared to No-RIS) increase with $N_i$, while the benefits of optimizing RIS elements decrease with $N_i$. It should be noted that the benefit of optimizing RIS components in more than $25\%$ for $N_i=32$ in this particular example, which is still significant.

\section{Conclusion}\label{sec-c}
%\vspace{-.1cm}
This paper proposed resource allocation algorithms with both precoding and RIS element optimizations for a multi-user MIMO RIS-aided OFDM BC with HWI to maximize the SE and EE. We showed that RIS can substantially enhance the SE and EE of the OFDM BC even when the number of RIS elements is  low. Moreover, we showed that the benefits of optimizing RIS elements in low SNR regimes are much higher than in high SNR regimes. Additionally, we showed that the benefits of optimizing RIS elements are still significant in OFDM systems even though the RIS coefficients cannot be optimized at each subband independently. 
Furthermore, we showed that IQI can significantly reduce the average minimum rate  even if it is compensated by employing IGS. Of course, the performance degradation by neglecting IQI in the design can be very significant and highly increases with the level of the imbalance at the transceivers. 
We, moreover, showed that multi-sector BD-RIS can significantly outperform a STAR-RIS as well as a regular RIS. 

\bibliographystyle{IEEEtran}
\bibliography{ref2}
\end{document}

%% file: r.tex
\definecolor{mycolor1}{rgb}{0.00000,0.44700,0.74100}%
\definecolor{mycolor2}{rgb}{0.85000,0.32500,0.09800}%
\definecolor{mycolor3}{rgb}{0.92900,0.69400,0.12500}%
\begin{tikzpicture}

\begin{axis}[%
width=6.5cm,
height=4.cm,
at={(2.311in,0.971in)},
scale only axis,
xmin=0,
xmax=17,
xlabel style={font=\color{white!15!black}},
xlabel={$P$ (dB)},
ymin=0,
ymax=35,
ylabel style={font=\color{white!15!black}},
ylabel={Minimum Rate (b/s/Hz)},
axis background/.style={fill=white},
axis x line*=bottom,
axis y line*=left,
xtick={0,4,8,12,17},
    grid=major,
ytick={10,20,30,35},
grid=major,
%ymajorgrids,
legend style={at={(0.01,0.56)}, anchor=south west, %legend columns=2, 
legend cell align=left, align=left, draw=white}
]
\addplot [color=blue, line width=1.5pt, mark=*, mark options={solid, fill=blue, blue}]
  table[row sep=crcr]{%
0	10.1879296704887\\
10	24.1140883211823\\
16.9897000433602	34.986776852281\\
};
\addlegendentry{RIS}

\addplot [color=red, line width=1.5pt, mark=triangle*, mark options={solid, rotate=90, fill=red, red}]
  table[row sep=crcr]{%
0	4.71434311475034\\
10	14.5248516023874\\
16.9897000433602	25.2964972304741\\
};
\addlegendentry{RIS-Rand}

\addplot [color=black, line width=1.5pt]
  table[row sep=crcr]{%
0	1.18860281233883\\
10	5.44002997030249\\
16.9897000433602	12.5410884268926\\
};
\addlegendentry{No-RIS}

\end{axis}
\end{tikzpicture}%

%% file: r-star.tex
\definecolor{mycolor1}{rgb}{0.00000,0.44700,0.74100}%
\definecolor{mycolor2}{rgb}{0.85000,0.32500,0.09800}%
\definecolor{mycolor3}{rgb}{0.92900,0.69400,0.12500}%
\definecolor{mycolor4}{rgb}{0.49400,0.18400,0.55600}%
\definecolor{mycolor5}{rgb}{0.46600,0.67400,0.18800}
\begin{tikzpicture}

\begin{axis}[%
width=6.5cm,
height=5.5cm,
at={(2.311in,0.971in)},
scale only axis,
xmin=0,
xmax=17,
xlabel style={font=\color{white!15!black}},
xlabel={$P$ (dB)},
ymin=0,
ymax=120,
ylabel style={font=\color{white!15!black}},
ylabel={Minimum Rate (b/s/Hz)},
axis background/.style={fill=white},
axis x line*=bottom,
axis y line*=left,
xtick={0,4,8,12,17},
    grid=major,
ymajorgrids,
legend style={at={(0.03,0.72)}, anchor=south west, legend columns=2, legend cell align=left, align=left, draw=white}
]
\addplot [color=blue, line width=1.5pt, mark=*, mark options={solid, fill=blue, blue}]
  table[row sep=crcr]{%
0	26.9503\\
10	68.1073\\
16.9897000433602	101.6433\\
};
\addlegendentry{S-RIS-ES}

\addplot [color=red, line width=1.5pt, dashed]
  table[row sep=crcr]{%
0	25.9559\\
10	64.5095\\
16.9897000433602	94.5812\\
};
\addlegendentry{S-RIS-ES$_N$\!\!}

\addplot [color=black, line width=1.5pt, mark=*, dash pattern={on 6pt off 2pt on 2pt off 3pt}, mark options={solid, fill=black, black}]
  table[row sep=crcr]{%
0	21.6313235333437\\
10	58.676644345135\\
16.9897000433602	91.7393534774092\\
};
\addlegendentry{S-RIS-MS}

\addplot [color=red, line width=1.5pt, dash pattern={on 3pt off 1pt on 1pt off 3pt}, mark=triangle*, mark options={solid, rotate=90, fill=red, red}]
  table[row sep=crcr]{%
0	11.8183618841706\\
10	37.6970175674958\\
16.9897000433602	68.4664710240467\\
};
\addlegendentry{S-RIS-Rand\!\!\!\!}

\addplot [color=black, dash pattern={on 2pt off 1pt on 4pt off 3pt}, line width=1.5pt]
  table[row sep=crcr]{%
0	6.56644046167047\\
10	28.9728525143703\\
16.9897000433602	59.5970466701349\\
};
\addlegendentry{RIS}

\addplot [color=mycolor5,  dash pattern={on 7pt off 2pt on 1pt off 3pt}, line width=1.pt, mark=diamond*, mark options={solid, rotate=90, fill=mycolor5,mycolor5}]
  table[row sep=crcr]{%
0	3.87802161610443\\
10	18.3945488671632\\
16.9897000433602	41.4292972487794\\
};
\addlegendentry{No-RIS}

\end{axis}
\end{tikzpicture}%